\documentclass[12pt]{article}
\usepackage{a4wide}
\usepackage{graphicx}
\newcommand{\be}{\begin{equation}}
\newcommand{\ee}{\end{equation}}
\newcommand{\ba}{\begin{eqnarray}}
\newcommand{\ea}{\end{eqnarray}}
\begin{document}
\begin{titlepage}
\begin{flushright}
\end{flushright}
\vfill
\begin{center}
{\Large {\bf Chiral and volume extrapolation of pion and
kaon electromagnetic form factor within SU(3) ChPT}} \\
\vspace{1cm}
{\bf Karim Ghorbani\footnote{email: kghorbani@ipm.ir} } \\[1cm]
\end{center}
{Physics Department, Faculty of Sciences, Arak University, Arak 38156-8-8349, Iran}
\vspace{2.7cm}
\begin{abstract}
We calculate the pion and kaon electromagnetic form factors
in finite volume for a generic momentum transfer in
space-like region, using SU(3) chiral perturbation theory.
To this end, we first find the hadronic matrix element
in a new form which is suitable for our calculation in
finite volume. We present our numerical findings for
the chiral behavior and finite volume corrections
and compare these for pion and kaon form factors.
As a result, for the pion electromagnetic form factor
we find the finite volume correction
$\Delta F^{\pi}_{em} = 0.0041$ at momentum transfer
$q^2 = -0.1358$ and for the kaon form factor $\Delta F^{K}_{em} = 0.0029$
at $q^2 = -0.1357$ with $m_{\pi} = 0.3128$ and
$L = 2.6 fm$ as the linear size of the lattice.
\end{abstract}

\vspace{3.5cm}

\end{titlepage}

\section{Introduction}
\label{intro}
Pion electromagnetic form factor is an interesting
low energy observable to be evaluated
in lattice QCD. Two early original works on
pion form factor calculations in lattice QCD
are pioneered in \cite{Martinelli1988,Draper1989}.
Further investigations in this direction
are pursued by other authors
in\cite{Heide2004,Abdel-Rehim2005,Capitani2006,JLQCD and TWQCD collaborations,Nguyen2011}.
In recent years, numerical calculation in lattice QCD
with relatively small quark masses and
larger finite volume has become feasible,
see for instance \cite{Bazavov:2009fk,Saoki2009,Saoki2010,DurrScience,Durr2010}.
The evaluation of this observable in lattice
QCD exemplifies an ideal case
due to the fact that here quark-disconnected
diagrams for two degenerate flavours are absent.
This, in turn, makes possible a precision calculation
of the pion vector form factor in lattice QCD.
Pion vector form factor is recently
studied in two-flavour
lattice QCD in a lattice with spacial linear
size of 1.83 [fm] with exact
chiral symmetry by authors in \cite{JLQCD and TWQCD collaborations}.
Moreover, recent results on the measurement of the electromagnetic form factor
of pion in Lattice QCD with dynamical quarks exist in \cite{Nguyen2011}.
Lattice data, however, comes along with
some important side effects caused
by the finite lattice size, large quark masses and lattice spacing.
Much efforts are spent on improving
algorithms to reach smaller quark masses and larger lattice size,
see, e.g. \cite{Bazavov:2009fk,Saoki2009,Saoki2010,DurrScience,Durr2010},
but it still remains to reach the physical condition.
To this end, effective field theories can be
exploited to make connection between lattice
data and the real world of QCD and thereby to experiment.
One another alternative to make this connection
is the L\"uscher approach to which we refer to
\cite{Luscher1,Luscher2} for detail discussion.

In this article we apply chiral perturbation theory (ChPT) with
three flavours in the mesonic sector to study chiral and volume dependence
of the electromagnetic form factors.
Within SU(3) ChPT, this is the first work that finite volume effects
are studied at arbitrary momentum transfer.
Although, the application
of ChPT in such studies has become a standard practice, one can still
ask how far we can push the realm of its applicability to a process
incorporating external current. Thus, one principal motivation
to undertake the current research is to find out an answer to the question
we usually do not know its answer beforehand.
As an instance in this regard, in a recent work we applied SU(3) ChPT to study
chiral extrapolation and finite volume effects of the semileptonic
kaon scalar form factor at the maximum momentum transfer \cite{Kgh2011}.
We line up, here, earlier works done in the framework of ChPT or ChPT in
combination with L\"uscher method.

Finite volume effects in pion mass are widely studied in \cite{Colangelo2003,Colangelo2005,Colangelo2006}
and for the pion decay constant in \cite{Colangelo2005,Colangelo2004}.
For a review discussion on the application of chiral perturbation theory
in finite volume, see \cite{Colangelo-review}.
Following the same line of reasoning, quark vacuum expectation value
is calculated in a finite box \cite{Bijnens-Ghorbani2006}.
There is also a calculation in the framework of partially quenched
chiral perturbation theory on finite volume effects
of the pion charge radius \cite{Bunton}.
In addition, pion pion scattering parameters are evaluated in finite
volume in reference \cite{Bedaque2006}.

The organization of this article is as follows.
First, we introduce chiral perturbation theory in
infinite and finite volume. On Sec.\,\ref{definition}
pion and kaon electromagnetic form factors are defined.
Our analytical results in its new form are provided
by Sec.\,\ref{analytical}.
Feynman integrals in finite volume are evaluated
on Sec.\,\ref{FVcalculation}.
Chiral extrapolation and finite volume effects
are studied on Sec.\,\ref{chiralextrapolation}
and Sec.\,\ref{formfactor-fv} respectively.
Finally, we finish up with conclusion on Sec.\,\ref{summary}.

\section{ChPT in infinite and finite volume}
\label{chpt}
\subsection{ChPT in infinite volume}
Quantum Chromo Dynamics (QCD) is strongly coupled at energies below the proton mass.
Therefore the standard perturbative technique is not a useful approach to calculate QCD
observables at low energy. Chiral perturbation theory (ChPT) is an effective field theory
which is proven successful in describing mesonic low energy QCD processes. The effective
Lagrangian organized in a series of operators as
\begin{equation}
\label{lagL}
\mathcal{L}_{eff} = \mathcal{L}_{2} + \mathcal{L}_{4} + \mathcal{L}_{6}
 + \cdots
\,.
\end{equation}
The subscribes show the chiral order. The expansion parameter is in
terms of external momenta $"p"$ and quark masses, $"m_{q}"$.
The lowest order SU(3) chiral Lagrangian involving two terms
has the following form, see \cite{Weinberg0}
\be
\label{Lp2}
{\cal L}_{2} = \frac{F_{0}^{2}}{4} \langle u_{\mu} u^{\mu}+ \chi_{+} \rangle,
\ee
where $F_{0}$ is the pion decay constant at chiral limit.
The notation $\langle...\rangle$ = $ \mathrm{Tr}_F\left(...\right)$
indicates the trace over the flavors.
we define the matrices $u^{\mu}$ and $\chi_{\pm}$ as following
\ba
u_{\mu} = i u^{\dag}D_{\mu}U u^{\dag} = u_{\mu}^{\dag}  \,, \quad u^{2} = U,
\nonumber\\
\chi_{\pm} = u^{\dag} \chi u^{\dag} \pm u\chi^{\dag}u.
\ea
The matrix $U \in SU(3)$ contains
the octet of the light pseudo-scalar mesons with its exponential
representation given in terms of meson fields
matrix as
\be
U(\phi) = \exp(i \sqrt{2} \phi/F_0)\,,
\ee
where
\ba
\phi (x)
 = \, \left( \begin{array}{ccc}
\displaystyle\frac{ \pi_3}{ \sqrt 2} \, + \, \frac{ \eta_8}{ \sqrt 6}
 & \pi^+ & K^+ \\
\pi^- &\displaystyle - \frac{\pi_3}{\sqrt 2} \, + \, \frac{ \eta_8}
{\sqrt 6}    & K^0 \\
K^- & \bar K^0 &\displaystyle - \frac{ 2 \, \eta_8}{\sqrt 6}
\end{array}  \right) .
\ea
Quark masses are counted of order $p^2$ due to the lowest
order mass relation $m_{\pi}^2 = B_{0}(m_{u}+m_{d})$.
The external fields are defined through the covariant derivatives as
\ba
D_{\mu} U = \partial_{\mu} U - i r_{\mu}U +iUl_{\mu}.
\ea
The right-handed and left-handed external fields are expressed
by $r_{\mu}$ and $l_{\mu}$ respectively. In the present work we set
\ba
r_{\mu} = l_{\mu}
 = e~A_{\mu} \left( \begin{array}{ccc}
\displaystyle 2/3 &   \\
    &\displaystyle -1/3 &  \\
 &   &\displaystyle -1/3
\end{array}  \right) .
\ea
The electron charge is denoted by $e$ and $A_{\mu}$ is the classical photon field.
The Hermitian $3\times3$
matrix $\chi$ involves the scalar (s) and pseudo-scalar external
densities and is given by $\chi = 2B_{0}(s+ip)$.
The constant $B_{0}$ is related to the pion decay
constant and quark condensate. For our purpose it suffices to write
\ba
\chi
 = 2B_{0}\, \left( \begin{array}{ccc}
\displaystyle m_{u} &   \\
    &\displaystyle m_{d} &  \\
 &   &\displaystyle m_{s}
\end{array}  \right) .
\ea

The next to leading order Lagrangian consists of 10+2 independent operators
with corresponding effective low energy constants (LEC's)\cite{GL0,Cirigliano2002,GL1}
\begin{eqnarray}
\label{lagL4}
{\cal L}_4&&\hspace{-0.5cm} =
L_1 \langle u_{\mu} u^{\mu} \rangle^2
+L_{2} \langle u_{\mu} u^{\nu} \rangle \langle u^{\mu} u_{\nu} \rangle
+L_{3} \langle u_{\mu} u^{\mu} u_{\nu} u^{\nu} \rangle
+L_{4} \langle u_{\mu} u^{\mu} \rangle \langle \chi_{+} \rangle
\nonumber\\&&\hspace{-0.1cm}
+L_{5} \langle u_{\mu} u^{\mu} \chi_{+} \rangle
+L_{6} \langle \chi_{+} \rangle^2+ L_{7} \langle \chi_{-} \rangle^2
+\frac{1}{4}(2L_{8}+L_{12})\langle \chi_{+}^2 \rangle
\nonumber\\&&\hspace{-0.1cm}
+\frac{1}{4}(2L_{8}-L_{12})\langle \chi_{-}^2 \rangle
-iL_{9} \langle f^{\mu \nu}_{+} u_{\mu} u_{\nu} \rangle
+\frac{1}{4}(L_{10}+2L_{11}) \langle f_{+\mu \nu} f_{+}^{\mu \nu}  \rangle
\nonumber\\&&\hspace{-0.1cm}
-\frac{1}{4}(L_{10}-2L_{11}) \langle f_{-\mu \nu} f_{-}^{\mu \nu}  \rangle.
\end{eqnarray}
The field strength tensor is defined as
\begin{eqnarray}
f_{\pm}^{\mu \nu}&&\hspace{-0.5cm} = u F^{\mu \nu}_{L} u^{\dag} \pm u^{\dag} F_{R}^{\mu \nu} u,
\nonumber\\
F^{\mu \nu}_{L}&&\hspace{-0.5cm} = \partial^{\mu}l^{\nu}-\partial^{\nu}l^{\mu}-i[l^{\mu},l^{\nu}],
\nonumber\\
F^{\mu \nu}_{R}&&\hspace{-0.5cm} = \partial^{\mu}r^{\nu}-\partial^{\nu}r^{\mu}-i[r^{\mu},r^{\nu}].
\end{eqnarray}

\subsection{ChPT application in finite volume}
\label{chpt-application}
Chiral perturbation theory can be applied to study
large volume effects which manifest
themselves on the radiative corrections.
Original works introducing the application
are provided by Gasser and Leutwyler in
\cite{Gasser1987finite1,Gasser1987finite2,Gasser1988finite3}.
There we can see that when particle fields are subjected to periodic
boundary condition, their momentum becomes discrete.
The momentum quantization leads to the modification of the
quantum corrections.
To make sure that the physics we study in finite volume is
almost the same as one in infinite volume, the volume size has
to be large enough, i.e. $m_{\pi} L >> 1$. This is the so called
{\it p-regime} and we do our study in this regime.
Besides, the applicability of ChPT is restricted
to particle momentum smaller than the chiral symmetry breaking
scale. In terms of the spacial size of the box, this corresponds
to the following condition
\begin{equation}
L >> \frac{1}{2F_{\pi}} \,.
\end{equation}

\section{The definition of the vector form factor}
\label{definition}
Lorentz invariance restricts the general structure of the
pseudoscalar-pseudoscalar electromagnetic form factor.
One can then define the pion and kaon matrix elements
by additional use of the charge conjugation and electromagnetic
gauge symmetry as
\begin{eqnarray}
<\pi^{+}(p^\prime)|j_{\mu}|\pi^{+}(p)> = (p_{\mu}+p_{\mu}^\prime) F^{\pi^+}_{V}(t) \,,
\end{eqnarray}
\begin{eqnarray}
<K^{+}(p^\prime)|j_{\mu}|K^{+}(p)> = (p_{\mu}+p_{\mu}^\prime) F^{K^+}_{V}(t) \,,
\end{eqnarray}
with $t = (p^\prime-p)^2$.
The current $j_{\mu}$ stands for the electromagnetic
current of quarks. The quantities $F^{\pi^+}_{V}(t)$
and $F^{K^+}_{V}(t)$ define the pion and kaon electromagnetic
form factors respectively, where we call them form factor hereafter.
The electromagnetic current $j_{\mu}$ is approximated by
the current due to the light flavours and is given by
\begin{eqnarray}
j_{\mu} = \frac{2}{3}(\bar u \gamma_{\mu} u) - \frac{1}{3}(\bar d \gamma_{\mu} d + \bar s \gamma_{\mu} s).
\end{eqnarray}

\section{Analytical results}
\label{analytical}
\begin{figure}
\begin{center}
\includegraphics[width=0.9\textwidth]{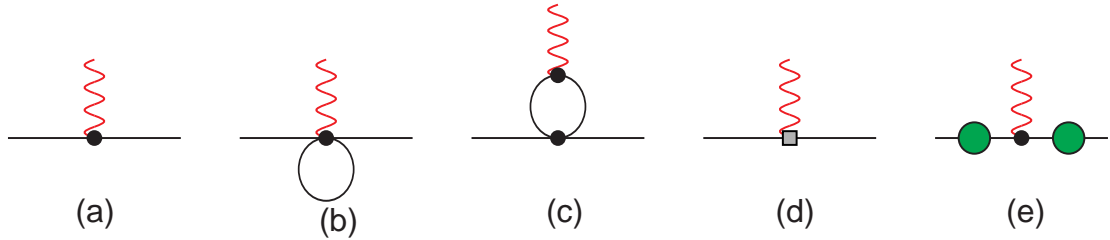}
\caption{Feynman diagrams contributing to the form factors are shown.
Diagram (a) is of ${\cal O}(p^2)$ with a vertex from the Lagrangian ${\cal L}_{2}$.
Other diagrams are of ${\cal O}{(p^4)}$.
Diagram (d) is a tree diagram with a vertex from the Lagrangian ${\cal L}_{4}$
and diagram (e) is the wave function renormalization of the external legs.
The thick lines stand for meson propagations and external photons
are denoted by curly lines.}
\label{diagrams}
\end{center}
\end{figure}

Pion and kaon matrix elements in chiral limit are recalculated
in this section up to order $p^4$.
In addition, we present the matrix elements
in a new form which can be used later on to
study form factors in finite volume.
We also recover the known expressions for the form factors
in infinite space. In Fig. \ref{diagrams} all relevant
Feynman diagrams including wave function renormalization
diagram are depicted.
\vspace{.5cm}
\subsection{Form factors at ${\cal O}(p^2)$ and ${\cal O}(p^4)$ }

At the leading order, one calculates the tree Feynman diagram shown
in Fig.\ref{diagrams}(a). In the isospin limit pion and kaon form
factors at order $p^2$ are
\begin{eqnarray}
F^{\pi^{+}}_{em}(t)|_{p^2} = F^{K^{+}}_{em}(t)|_{p^2} = 1.
\end{eqnarray}
At the next-to-leading order, there are four contributing Feynman diagrams
to the matrix element which are depicted in Fig.\ref{diagrams}(b-e).
We evaluate each of the Feynman diagrams separately
and find for sum of all the diagrams the following
results in terms of tensor integrals
\begin{eqnarray}
\label{ptensor}
F_{em}^{\pi^{+}}(p,p^{\prime}).\epsilon &&\hspace{-0.5cm} =
\frac{1}{F_{\pi}^2} \Big [ [2 q^{2} L_{9} + A(m_{\pi}^2)+\frac{1}{2}A(m_{K}^2)] r.\epsilon
+\frac{1}{3}A(m_{\pi}^2)q.\epsilon
\nonumber\\&&\hspace{-0.5cm}
+B(m_{\pi}^2,m_{\pi}^2,q^2)\Big [\frac{1}{3}q^2- m_{\pi}^2 \Big ]q.\epsilon
-B_{\mu \nu}(m_{\pi}^2,m_{\pi}^2,q^2)\Big[2r^{\mu} \epsilon^{\nu} +
\frac{2}{3} q^{\mu} \epsilon^{\nu} \Big]
\nonumber\\&&\hspace{-0.5cm}
-B_{\mu \nu}(m_{K}^2,m_{K}^2,q^2) r^{\mu} \epsilon^{\nu}
+B_{\mu}(m_{\pi}^2,m_{\pi}^2,q^2)\Big[r^{\mu} q.\epsilon+\frac{1}{3}q^{\mu} q.\epsilon-
\frac{2}{3} q^2 \epsilon^{\mu}+2m_{\pi}^2 \epsilon^{\mu} \Big]
\nonumber\\&&\hspace{-0.5cm}
+\frac{1}{2}B_{\mu}(m_{K}^2,m_{K}^2,q^2) r^{\mu} q.\epsilon \Big ]\,
\end{eqnarray}
and
\begin{eqnarray}
\label{ktensor}
F_{em}^{K^+}(p,p^{\prime}).\epsilon &&\hspace{-0.5cm} =
\frac{1}{F_{\pi}^2} \Big [ [2 q^{2} L_{9} + \frac{1}{2} A(m_{\pi}^2)+ A(m_{K}^2)] r.\epsilon
+\frac{1}{3}A(m_{K}^2)q.\epsilon
\nonumber\\&&\hspace{-0.5cm}
+ B(m_{K}^2,m_{K}^2,q^2)\Big [\frac{1}{3} q^2- m_{K}^2 \Big ] q.\epsilon
-B_{\mu \nu}(m_{K}^2,m_{K}^2,q^2)\Big[2r^{\mu} \epsilon^{\nu} + \frac{2}{3} q^{\mu} \epsilon^{\nu} \Big]
\nonumber\\&&\hspace{-0.5cm}
-B_{\mu \nu}(m_{\pi}^2,m_{\pi}^2,q^2) r^{\mu} \epsilon^{\nu}
+B_{\mu}(m_{K}^2,m_{K}^2,q^2)\Big[r^{\mu} q.\epsilon+\frac{1}{3}q^{\mu} q.\epsilon-
\frac{2}{3} q^2 \epsilon^{\mu}+2m_{K}^2 \epsilon^{\mu} \Big]
\nonumber\\&&\hspace{-0.5cm}
+\frac{1}{2}B_{\mu}(m_{\pi}^2,m_{\pi}^2,q^2) r^{\mu} q.\epsilon \Big ]\,
\end{eqnarray}
with
\begin{eqnarray}
r = p^{\prime}+p \,,  &&\hspace{.5cm}  q = p^{\prime} -p.
\end{eqnarray}
Here, $\epsilon$ is the polarization four vector of the photon.
The integrals $A$ and $B$ along with the tensor integrals
$B_{\mu}$ and $B_{\mu \nu}$ are defined in Appendix A.
These tensor integrals can be written in terms of scalar
functions. The relevant relations known as reduction formulas
are given in Appendix A.
Scalar one-loop integrals are evaluated in the
dimensional regularization scheme.
To absorb the infinities arising from
the integrals, the LEC's are redefined
in terms of the renormalized LEC's and
subtracted infinities as discussed in \cite{GL1}.
All infinities cancel out and we obtain the known finite
results for the pion form factor

\begin{eqnarray}
\label{pionform}
F_{em}^{\pi^{+}}(t) &=&
\frac{1}{F_\pi^2}\Big[ 2\,L^r_{9}t+ \overline{A}(m_{\pi}^2)+
\frac{1}{2}\,\overline{A}(m_{K}^2)
\nonumber\\&&
         - 2\,\overline{B}_{22}(m_{\pi}^2,m_{\pi}^2,t)-\overline{B}_{22}(m_{K}^2,m_{K}^2,t)\Big],
\end{eqnarray}
and for the kaon form factor
\begin{eqnarray}
\label{kaonform}
F_{em}^{K^+}(t) &=&
\frac{1}{F_\pi^2}\Big[ 2\,L^r_{9}t+\frac{1}{2}\, \overline{A}(m_{\pi}^2)+
\overline{A}(m_{K}^2)
\nonumber\\&&
         - \overline{B}_{22}(m_{\pi}^2,m_{\pi}^2,t)-2\,\overline{B}_{22}(m_{K}^2,m_{K}^2,t)\Big].
\end{eqnarray}

\section{Finite Volume Calculations}
\label{FVcalculation}
To study the volume effects, we obtain vector
form factors from temporal component
of the meson-meson matrix element while it is
also possible to extract these
quantities form spatial components.
Vector form factor of a charged pseudoscalar meson, M,
in finite volume is then defined by
\begin{eqnarray}
F^{M}_{finite}  = \frac{<M(p^\prime)|j_{0}|M(p)>_{V}}{E_{\vec p}+E_{\vec p'}}.
\end{eqnarray}
In this section we evaluate necessary one-loop Feynman
integrals at momentum transfer $q^2$ which appear in the
temporal component of the expressions presented
in Eq.~\ref{ptensor} and Eq.~\ref{ktensor}.
In the following we define for a generic
function $G$, $\Delta G = G_{V} - G_{\infty}$.
Here subscripts $V$ and $\infty$ stand for integration
in finite and infinite volume, respectively.
The simplest integral we encounter in our expression
for the matrix element is the tadpole integral
for which we have
\begin{eqnarray}
\Delta A = - \frac{m}{4 \pi^2 L} \sum_{\vec{n} = \vec{1}}
\frac{1}{\vert \vec{n} \vert} m(k) K_{1}(m L \vert \vec{n}\vert),
\end{eqnarray}
where, $K_{1}$ is the modified Bessel function of order one
and the coefficient $m(k)$ accounts for the number of
possibilities that the relation $k = n_{1}^2+n_{2}^2+n_{3}^2$
holds. We drop $m(k)$ in our formulas hereafter for the sake of simplicity
but it is taken into account in numerical evaluations.
The first new Feynman integral we need to evaluate in
finite volume is
\begin{eqnarray}
B(m^2,m^2,q^2) =  \frac{1}{i} \int \frac{d^dp}{(2\pi)^d}
\frac{1}{(p^2-m^2)((q+p)^2-m^2)}.
\end{eqnarray}
When we put a system in a finite volume, momentum $p$ gets
discrete values $\vec p = \frac{2\pi}{L} \vec n $,
therefore we should calculate the integral
\begin{eqnarray}
\label{bfunc}
B_{V} (m^2,m^2,q^2)  = -\frac{i}{L^3} \sum_{\vec{p}} \int \frac{dp_0}{2\pi}
\frac{1}{(p^2-m^2)((p+q)^2-m^2)}
\nonumber\\&&\hspace{-8.1cm}
= B_{\infty}-\int\frac{dp_0}{2\pi}\sum_{\vec{n}\neq 0}\int \frac{d^{3}\vec{p}}{(2\pi)^3}
                \frac{i e^{iL\vec{p}.\vec{n}}}{(p^2-m^2)((p+q)^2-m^2)}.
                \nonumber \\
\end{eqnarray}
To get the second line in the equation above we used the
Poisson summation formula
\begin{eqnarray}
\frac{1}{L^3} \sum_{\vec{p}=\frac{2\pi}{L} \vec{n}} f(\vec{p}~^2) = \int \frac{d^3p}{(2\pi)^3} f(\vec{p}~)
+\sum_{\vec{n}\neq 0} \int \frac{d^3p}{(2\pi)^3} f{(\vec p~)} e^{iL\vec{p}.\vec{n}}.
\end{eqnarray}
Making use of the Feynman parameter formula in Eq.\ref{bfunc} followed by
redefining the variable $p_{0}$ will get us at
\begin{eqnarray}
\Delta B (m^2,m^2,q^2) =
-i \int_{0}^{1}dx \int \frac{dp_0}{2\pi} \sum_{\vec{n}\neq 0}
\int \frac{d^{3}\vec{p}}{(2\pi)^3}
\frac{e^{iL\vec{p}.\vec{n}}}{\Large[ p_{0}^2-(\vec{p}+(1-x)\vec{q}~)^2+x(1-x)q^2-m^2 \Large]^2}.
\end{eqnarray}
At this stage, we begin by taking the contour integral over $p_{0}$
and then follow up by redefining the variable $\vec{p}$ to find
\begin{eqnarray}
\Delta B (m^2,m^2,q^2) =
\frac{1}{4}\sum_{\vec{n}\neq 0} \int_{0}^{1} dx~e^{-iLx\vec{q}.\vec{n}}
\int \frac{d^{3}\vec{p}}{(2\pi)^3}
\frac{e^{iL\vec{p}.\vec{n}}}{[\vec{p}^2+m^2-x(1-x)q^2]^{3/2}}.
\end{eqnarray}
It is evident that the exponential factor $e^{-iLx\vec{q}.\vec{n}}$
explicitly breaks the rotational symmetry.
We carry out the integral over the vector
momentum in two steps. First we take an integral
over the angular part of the three dimensional momentum.
We then make use of the convolution technique
to perform the last integral. We achieve the following result
\begin{eqnarray}
\label{Bfunc}
\Delta B(m^2,m^2,q^2) =  \frac{1}{8\pi^2} \sum_{\vec{n} \neq 0}
\int_{0}^{1} dx~ C_{n}(\alpha x)~K_{0}(w~\sqrt{m^2-x(1-x)q^2})\,,
\end{eqnarray}
where $w = L |\vec{n}|$. $K_{0}$ is the modified Bessel function of rank one.
Functions $C_{n}(\alpha x)$ introduced
above incorporate the exponential factor when we sum over all
possible ways that for a given $n$ the relation
$n = n_{1}^2+n_{2}^2+n_{3}^2$ holds. For instance, in the case that
we assume the momentum transfer, $\vec{q}$, has equal components, i.e,
$q_{x}=q_{y}=q_{z}=q_{s}$, we obtain functions $C_{n}(\alpha x)$
which are provided by Appendix B, where $\alpha = L q_{s}$.
A discussion on various choices for a given external momentum transfer
in estimating finite volume effects of the nucleon magnetic moment
can be found in \cite{Tiburzi2007}.
For a special case where $q = 0$, the integral above can be evaluated
analytically since we now have $C_{k}(\alpha x) = m(k)$ and therefore
\begin{eqnarray}
\label{Bfunc2}
\Delta B(m^2,m^2,0) = \frac{1}{8 \pi^2}  \sum_{\vec{n} \neq 0}
m(k)~K_{0}(m L \vert \vec{n} \vert).
\end{eqnarray}
This agrees with the formula one can obtain directly by setting $q=0$
in Eq.~\ref{bfunc}.
The next integral we need to consider in finite volume is
\begin{eqnarray}
B^{0}(m^2,m^2,q^2) =  \frac{1}{i} \int \frac{d^dp}{(2\pi)^d}
\frac{p_{0}}{(p^2-m^2)((q+p)^2-m^2)}.
\end{eqnarray}
$B^{0}$ is the temporal component of the tensor integral $B^{\mu}$
defined in Appendix A. By repeating the procedures
stated above we can readily prove that
\begin{eqnarray}
\label{b0func}
\Delta B^{0}(m^2,m^2,q^2) = -\frac{q_{0}}{8\pi^2} \sum_{\vec{n} \neq 0}
\int_{0}^{1} dx~x~C_{n}(\alpha x)~K_{0}(w~\sqrt{m^2-x(1-x)q^2})\,,
\end{eqnarray}
where, $q_{0} = E_{p^\prime}-E_{p}$.

The last Feynman integral to be evaluated in finite volume reads
\begin{eqnarray}
\label{b00integral}
B^{00}(m^2,m^2,q^2) =  \frac{1}{i} \int \frac{d^dp}{(2\pi)^d}
\frac{p_0^2}{(p^2-m^2)((q+p)^2-m^2)}.
\end{eqnarray}
Following the procedures sketched above we find that
\begin{eqnarray}
\label{b00func}
\Delta B^{00}(m^2,m^2,q^2)  =  -\frac{1}{8\pi^2 L} \sum_{\vec{n} \neq 0}
\frac{1}{|\vec n|} \int_{0}^{1} dx
~C_{n}(\alpha x)~\sqrt{m^2-x(1-x)q^2}\times
\nonumber\\&&\hspace{-5.8cm}
K_{1}(w~\sqrt{m^2-x(1-x)q^2})
\nonumber\\
+\frac{q_{0}^2}{8\pi^2 L} \sum_{\vec{n} \neq 0}
\frac{1}{|\vec n|} \int_{0}^{1} dx~
x^2~C_{n}(\alpha x)~K_{0}(w~\sqrt{m^2-x(1-x)q^2}).
\end{eqnarray}
We evaluate $\Delta B^{00}(m^2,m^2,0)$ directly form the
start by putting $q = 0$ in Eq.~\ref{b00integral} and obtain
\begin{eqnarray}
\Delta B^{00}(m^2,m^2,0) = -\frac{m}{8 \pi^2 L}  \sum_{\vec{n} \neq 0}
\frac{m(k)}{ \vert \vec{n} \vert} K_{1}(m L \vert \vec{n} \vert),
\end{eqnarray}
which agrees with Eq.~\ref{b00func} at $q = 0$.

\section{Chiral extrapolation}
\label{chiralextrapolation}
The question of how pion and kaon vector form factors depend
on pion mass is addressed in this section. The application
of SU(3) ChPT can also help study their kaon mass dependency as well.
Our formulas given in Eq.~\ref{pionform} and Eq.~\ref{kaonform}
 for the vector form factors
in infinite volume incorporate pion decay constant and
a low energy constant,
where we set as input $F_{\pi} = 0.0924 \,$GeV and
$L^{r}_{9}\,(\mu =0.770 \,$GeV$) = 0.0094$, respectively.
For a through discussion on the value of the
latter constant one may consult \cite{Amoros2001} and references therein.

\begin{figure}
\begin{center}
\includegraphics[angle=-90,width=.7\textwidth]{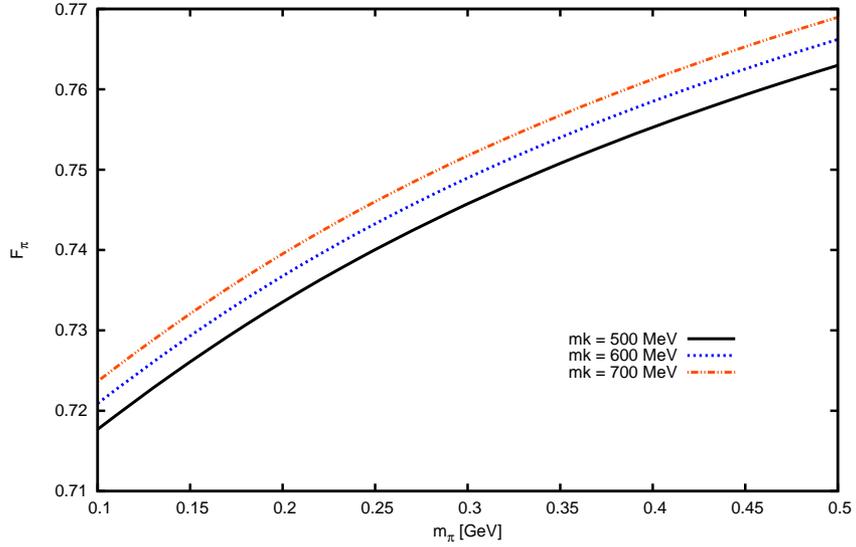}
\caption{Shown are the pion electromagnetic form factors
versus pion mass for three different values of kaon mass
at momentum transfer $t = -0.15\,$GeV$^2$.}
\label{infinity-pion}
\end{center}
\end{figure}

\begin{figure}
\begin{center}
\includegraphics[angle=-90,width=0.7\textwidth]{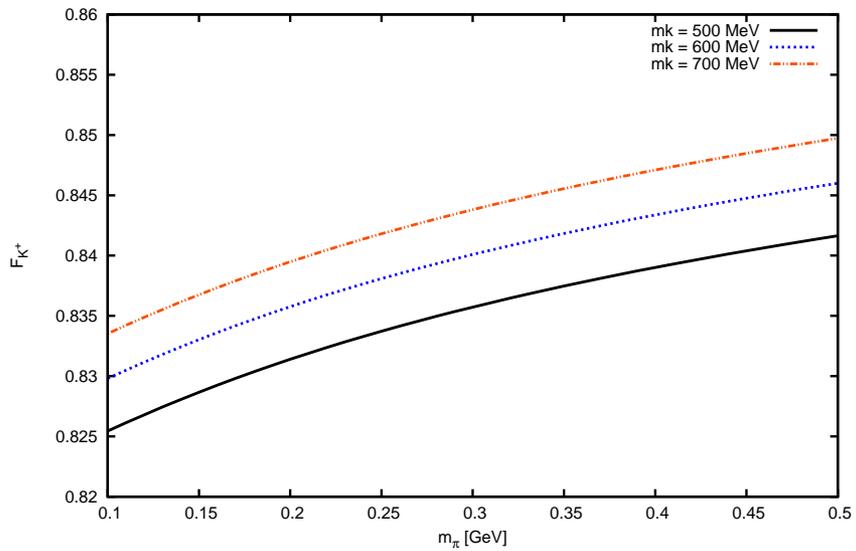}
\caption{Shown are the kaon electromagnetic form factors
versus pion mass for three different values of kaon mass
at momentum transfer $t = -0.10\,$GeV$^2$.}
\label{infy-kaon}
\end{center}
\end{figure}

Numerical results concerning these studies are displayed
in Fig.~\ref{infinity-pion} for pion form factor and in Fig.~\ref{infy-kaon}
for kaon form factor for three different values of kaon mass,
namely, $M_{K} = 500, 600, 700 \,$ MeV.
Momentum transfers in which these form factors are plotted
are $q^2 = -0.15\,$ GeV
for pion and $q^2 = -0.10\,$ GeV for kaon. As seen in the figures,
in the pion mass interval of $400\, $MeV we find about 5 percent
increase in the pion vector form factor and nearly 2 percent
enhancement in the kaon form factor.
\section{Form factors in finite volume}
\label{formfactor-fv}
In this section we first calculate the pion and kaon form factor
in finite volume for a set of input values as those quoted
in \cite{JLQCD and TWQCD collaborations} except that
we choose a bigger lattice size of $L =2.6\,fm$
rather than $L = 1.83\,fm$.
With $L = 2.6\,fm$ and $m_{\pi} = 0.3128\,$GeV
we have $m_{\pi} L = 4.06$ as large enough as it is required
to be confident that the condition $m_{\pi} L >> 1$
is fulfilled. Our numerical results for the pion form factor
at different values of momentum transfer are
summarized in Table.~\ref{tabl1}.
The size of the finite volume corrections are
about $0.5$ percent for pion form factor.
We perform the same type of analysis for the kaon
form factor at different values of momentum transfer
and provide our result in Table.~\ref{tabl2}.
As can be seen from Table.~\ref{tabl2} for kaon,
the size of the finite volume corrections are
smaller than that of pion and is nearly $0.25$ percent.
\begin{table}
\begin{center}
  \footnotesize{
   \begin{tabular}{|c|c|c|c|c|c|c|c|}
\hline
 $ \bf{ \vec{p} \,(GeV)}$ & $\bf{ \vec{p}' \,(GeV)} $  & $ \bf{E_{p} \, (GeV)} $ & $\bf{E_{p'} \, (GeV)} $ & $\bf{q^2 \, (GeV^2)} $
  & $\bf{\Delta F^{\pi}_{em} \, }$ & $ \bf{F^{\pi}_{\infty} } $  & $ \bf{F^{\pi}_{finite} }$  \\
\hline
0.3926   & 0      &   0.5019   & 0.3128 & -0.1183  & 0.0034    & 0.800 & 0.8034 \\
\hline
0.5552   & 0.3926  &   0.6372   & 0.5019  & -0.1358  & 0.0041  & 0.7707 & 0.7748 \\
\hline
0.6800   & 0.5552  &   0.7485   & 0.6372  & -0.1417  & 0.0044  & 0.7609 & 0.7653 \\
\hline
0.5552   & 0      &   0.6372   & 0.3128  & -0.2030  & 0.0036   & 0.6590  & 0.6626 \\
\hline
\end{tabular}
}
\caption{\label{tabl1}
Pion electromagnetic form factor in finite
volume in space-like region for various values of momentum
transfer $q^2$. As input we take $m_{\pi} = 0.3128\,$GeV, $m_{K} = 0.5\,$GeV
and the linear size of the lattice $L = 2.6~fm$ which
corresponds to $m_{\pi} L = 4.06$. First and second
column refer to vector momentum of final and initial
pion, respectively.}
\end{center}
\end{table}
\begin{table}
  \centering
\footnotesize{
   \begin{tabular}{|c|c|c|c|c|c|c|c|}
\hline
   $ \bf{ \vec{p} \,(GeV)}$ & $\bf{ \vec{p}' \,(GeV)} $  & $ \bf{E_{p} \, (GeV)} $ & $\bf{E_{p'} \, (GeV)} $ & $\bf{q^2 \, (GeV^2)} $
  & $\bf{\Delta F^{K}_{em} \, }$ & $ \bf{F^{K}_{\infty} } $  & $ \bf{F^{K}_{finite} }$  \\
\hline
  0.3926   & 0      &   0.6357   & 0.5 & -0.1357  & 0.0029   & 0.7781 & 0.7810 \\
  \hline
  0.5552   & 0.3926  &   0.7471   & 0.6357  & -0.1417 & 0.0028  & 0.7684 & 0.7712 \\
  \hline
  0.6800   & 0.5552  &   0.8440   & 0.7471  & -0.1447  &  0.0028 & 0.7636 & 0.7664 \\
  \hline
  0.5552   & 0      &   0.7471   & 0.5  & -0.2471      &  0.0035   & 0.5987  & 0.6022 \\
  \hline
\end{tabular}
}
\caption{\label{tabl2}
Kaon electromagnetic form factor in finite
volume in space-like region for various values of momentum
transfer $q^2$.
We take as input $m_{\pi} = 0.3128\,$GeV,
$m_{K} = 0.5\,$GeV and the linear size of
the lattice $L = 2.6~fm$ which corresponds
to $m_{\pi} L = 4.06$. First and second
column refer to vector momentum of final
and initial kaon, respectively.}
\end{table}

\begin{figure}
\begin{center}
\includegraphics[angle=-90,width=0.7\textwidth]{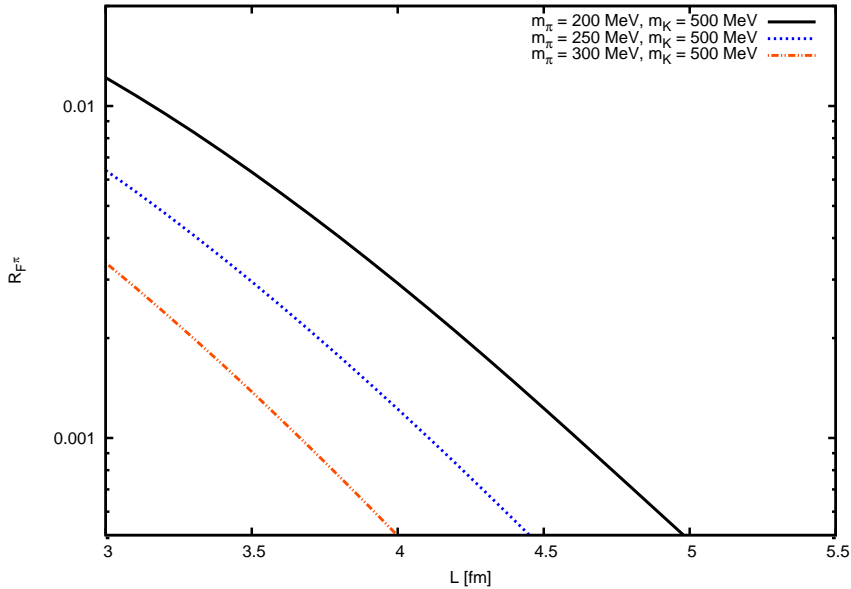}
\caption{The ratio $R_{F^{\pi}_{em}}(t)$ is plotted verses the linear size of the lattice
for three different mass of pion and a fixed mass of kaon at momentum transfer
t = -0.1358\, GeV$^2$.}
\label{finite-pion}
\end{center}
\end{figure}

\begin{figure}
\begin{center}
\includegraphics[angle=-90,width=0.7\textwidth]{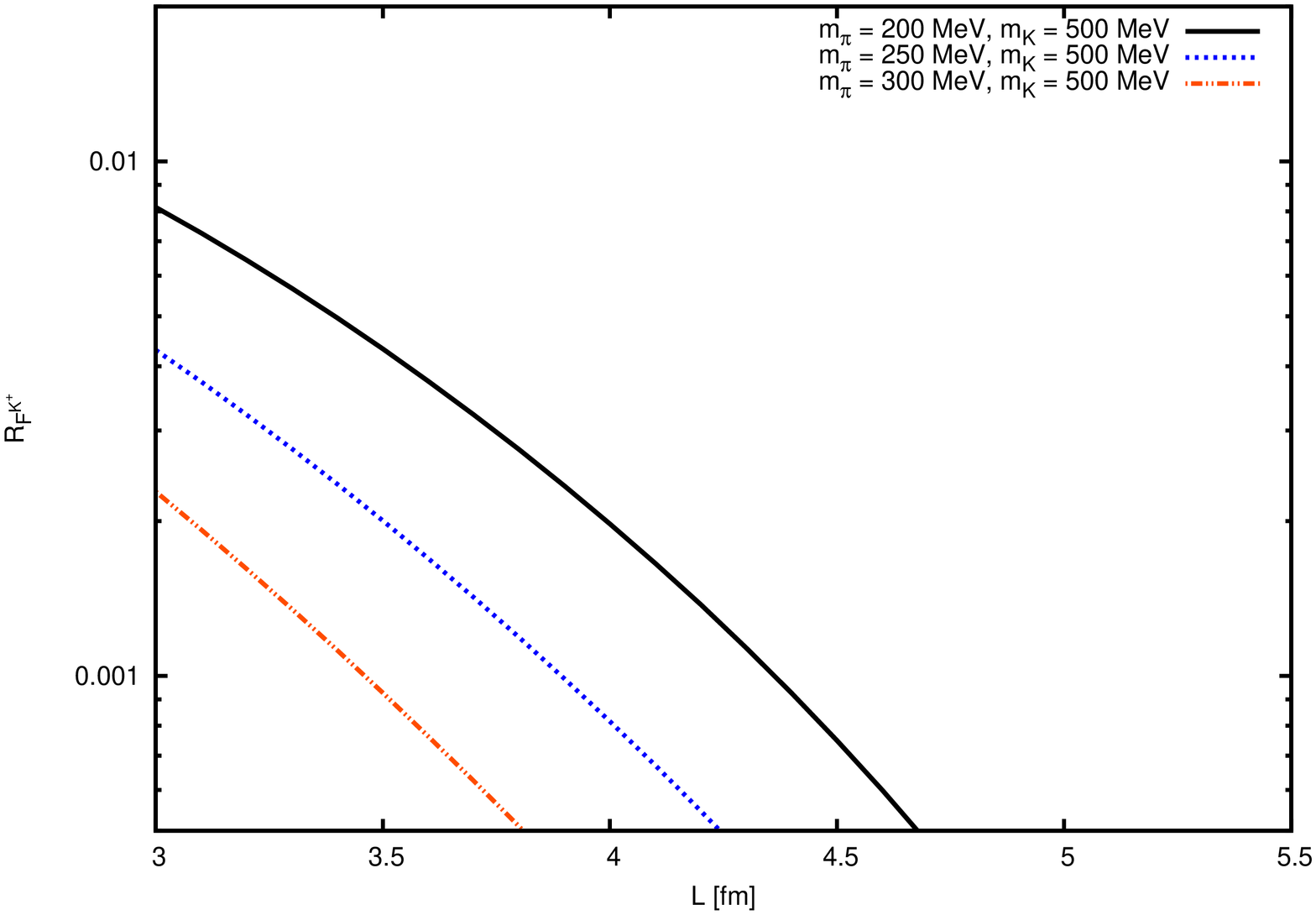}
\caption{The ratio $R_{F^{K}_{em}}(t)$ is plotted verses the linear size of the lattice
for three different mass of pion and a fixed mass of kaon at
momentum transfer t = -0.1357\, GeV$^2$.}
\label{finite-kaon}
\end{center}
\end{figure}

Moreover, we study the volume dependence of
the form factor at order $p^4$, given the following
definition of the ratio $R_{F^{M}_{em}}(t)$ for meson M
\begin{eqnarray}
R_{F^{M}_{em}}(t) =\frac{\Delta F^{M}_{em}(t)}{F^{M}_{\infty}(t)},
\end{eqnarray}
where $\Delta F^{M}_{em}(t) = F^{M}_{finite}(t)-F^{M}_{\infty}(t)$.
We keep the kaon mass fixed and evaluate the ratio numerically for
three different values of pion mass, namely, 200 MeV, 250 MeV and 300 MeV.
Our findings are shown in Fig.~\ref{finite-pion} for pion and
in Fig.~\ref{finite-kaon} for kaon form factor.
In obtaining the results we have kept the energy of the incoming
and the outgoing mesons fixed while we evaluate the ratio
versus lattice size for different values of pion mass.
We can get the standard observation out of
these graphs indicating that there is a larger volume
dependency of the form factors for larger pion mass
even at small volume. This is not actually the case for the
semileptonic kaon scalar form factor at the maximum momentum transfer
for small lattice size, see \cite{Kgh2011} to do comparison.

\section{Conclusion}
\label{summary}
We have put once again the applicability of
ChPT in finite volume under scrutiny.
We obtain the pion and kaon electromagnetic form
factor in finite volume at order ${\cal O}(p^4)$
for a generic momentum transfer.
We present some part of our result in the limit, $m_{\pi} L \approx 4$,
where ChPT application in finite volume is considered
trustworthy. In addition, we studied the volume dependence
of the form factor. Our results indicate a standard trend,
in the sense that finite volume corrections are large
for smaller pion mass.
This can be compared with the result in \cite{Kgh2011}
for the semileptonic kaon scalar form factor at the maximum
momentum transfer, in which for small volumes, finite volume
corrections become larger for larger pion mass.
It also turns out that finite volume correction of the
electromagnetic kaon form factor at any momentum
transfer is roughly two times smaller than that
of the pion form factor.

As an important step toward a further investigation
in this direction we have started a two-loop calculation
of the pion form factor in finite volume at a generic
momentum transfer in space-like region within SU(3) ChPT.
\section*{Acknowledgments}
I would like to thank Johan Bijnens and Hidenori Fukaya
for useful comments. I am specially very grateful to Brain Tiburzi
for his important comments. Arak University is kindly acknowledged
for financial support under the contract No.90/8479.
\section{Appendix A}
We introduce the necessary one loop Feynman integrals in what follows
\begin{eqnarray}
A(m^2) =  \frac{1}{i} \int \frac{d^dp}{(2\pi)^d} \frac{1}{p^2-m^2},
\end{eqnarray}
\begin{eqnarray}
B(m^2,m^2,q^2)&&\hspace{-0.1cm} =  \frac{1}{i} \int \frac{d^dp}{(2\pi)^d}
\frac{1}{(p^2-m^2)((p+q)^2-m^2)},
\end{eqnarray}
\begin{eqnarray}
B_{\mu}(m^2,m^2,q^2)&&\hspace{-0.1cm} =  \frac{1}{i} \int \frac{d^dp}{(2\pi)^d}
\frac{p_{\mu}}{(p^2-m^2)((p+q)^2-m^2)},
\end{eqnarray}
\begin{eqnarray}
B_{\mu \nu}(m^2,m^2,q^2)&&\hspace{-0.1cm} =  \frac{1}{i} \int \frac{d^dp}{(2\pi)^d}
\frac{p_{\mu} p_{\nu}}{(p^2-m^2)((p+q)^2-m^2)}.
\end{eqnarray}
By applying lorentz symmetry, we can write the tensor integrals in terms of scalar functions
\begin{eqnarray}
B_{\mu}(m^2,m^2,q^2) =  q_{\mu} B_{1}(m^2,m^2,q^2) ,
\end{eqnarray}
\begin{eqnarray}
B_{\mu \nu}(m^2,m^2,q^2)&&\hspace{-0.1cm} =  q_{\mu} q_{\nu} B_{12}(m^2,m^2,q^2)
+g_{\mu \nu} B_{22}(m^2,m^2,q^2).
\end{eqnarray}
\section{Appendix B}
In Table~\ref{functions}, we list functions $C_{n}(\alpha x)$ introduced in section \ref{FVcalculation},
given the assumption that the momentum transfer $\vec{q}$ has equal components.
\begin{table}
\begin{center}
 \footnotesize{
    \begin{tabular}{|c|c|c|c|}
\hline
n        &   $C_{n}(\alpha x)$   &  n    &  $C_{n}(\alpha x)$   \\
\hline
1       &   $6~cos(\alpha x)$    &  11   &  $6~cos(5\alpha x)+12~cos(3\alpha x)+6~cos(\alpha x)$                    \\
\hline
2       &   $6~cos(2\alpha x)+6$ &  12   &  $2~cos(6\alpha x)+6~cos(2\alpha x)$  \\
\hline
3       &   $2~cos(3\alpha x)+6~cos(\alpha x)$ & 13  &  $12~cos(5\alpha x)+12~cos(\alpha x)$    \\
\hline
4   &     $6~cos(2\alpha x)$  & 14  &  $12~cos(6\alpha x)+12~cos(4\alpha x)+12~cos(2\alpha x)+12$  \\
\hline
5   &     $12~cos(3\alpha x)+12~cos(\alpha x)$  & 15 & 0      \\
\hline
6   &     $6~cos(4\alpha x)+12~cos(2\alpha x)+6$  & 16 & $6~cos(4\alpha x)$   \\
\hline
7   &  0       & 17           &  $6~cos(7\alpha x)+12~cos(5\alpha x)+24~cos(3\alpha x)+6~cos(\alpha x)$    \\
\hline
8   &  $6~cos(4\alpha x)+6$  & 18  &  $12~cos(6\alpha x)+12~cos(4\alpha x)+6~cos(2\alpha x)+6$    \\
\hline
9   &  $6~cos(5\alpha x)+12~cos(3\alpha x)+12~cos(\alpha x)$  & 19 &  $8~cos(7\alpha x)+4~cos(5\alpha x)+12~cos(\alpha x)$     \\
\hline
10   &  $12~cos(4\alpha x)+12~cos(2\alpha x)$  & 20 & $12~cos(6\alpha x)+12~cos(2\alpha x)$     \\
\hline
\end{tabular}
}
\caption{\label{functions}
Functions $C_{n}(\alpha x)$ are provided for $1 \le n \le 20$ when external momentum $\vec{q} = q_{s}~(1,1,1)$.}
\end{center}
\end{table}

\end{document}